\documentclass[12pt,preprint]{aastex}
\usepackage{natbib}

\newcommand{\EQ}{\begin{equation}}
\newcommand{\EN}{\end{equation}}
\newcommand{\EQA}{\begin{eqnarray}}
\newcommand{\ENA}{\end{eqnarray}}
\newcommand{\UQ}{\begin{displaymath}}
\newcommand{\UN}{\end{displaymath}}

\newcommand{\NOCOMMENT}[1]{}

\begin{document}

\title{Dynamic Hydrogen Ionization}

\author{Mats Carlsson}
\affil{Institute of Theoretical Astrophysics, University of Oslo,
P.O. Box 1029 Blindern, N-0315 Oslo, Norway}
\email{Mats.Carlsson@astro.uio.no}

\and

\author{R. F. Stein}
\affil{Dept. of Physics and Astronomy, Michigan State University,
East Lansing, MI 48823, U.S.A.}
\email{stein@pa.msu.edu}

\begin{abstract}
We investigate the ionization of hydrogen in a dynamic Solar
atmosphere.  
%
The simulations include a detailed non-LTE treatment of hydrogen, calcium
and helium but lack other important elements. Furthermore, the omission of 
magnetic fields and the one-dimensional approach make the modeling
unrealistic in the upper chromosphere and higher. We discuss these
limitations and show that the main results remain valid for any 
reasonable chromospheric conditions.
As in the static case we find that the ionization of
hydrogen in the chromosphere is dominated by collisional excitation in
the Lyman-$\alpha$ transition followed by photoionization by Balmer
continuum photons --- the Lyman continuum does not play any
significant role. In the transition region, collisional ionization
from the ground state becomes the primary process.
%
We show that the time scale for ionization/recombination can be
estimated from the eigenvalues of a modified rate matrix where the
optically thick Lyman transitions that are in detailed balance have
been excluded.
We find that the time scale for ionization/recombination is dominated
by the slow collisional leakage from the ground state to the first
excited state. Throughout the chromosphere the time scale is long 
($10^3$-$10^5$ s), except in shocks where the increased temperature and density
shorten the time scale for ionization/recombination, especially in the
upper chromosphere. Because the relaxation time scale is much
longer than dynamic time scales, hydrogen ionization does not have time
to reach its equilibrium value and its fluctuations 
are much smaller than the variation of its statistical
equilibrium value appropriate for the instantaneous conditions. Because
the ionization and recombination rates increase with increasing
temperature and density, ionization in shocks is more rapid than
recombination behind them.  Therefore, the ionization state tends to
represent the higher temperature of the shocks, and the mean electron density
is up to a factor of six higher than the electron density calculated
in statistical equilibrium from the mean atmosphere.
The simulations show that a static picture and a dynamic picture
of the chromosphere are fundamentally different and that time
variations are crucial for our understanding of the chromosphere
itself and the spectral features formed there.
\end{abstract}



\section{Introduction}\label{sec:intro}

The solar chromosphere is the region about 1.5 Mm thick between the
temperature minimum (at about 4000 K) and the transition region to
coronal temperatures of several million Kelvin.  
Traditionally, its structure has been determined by 
semi-empirical fitting of temporally
and spatially averaged continua and line intensities
\citep[e.g.,][]{VALIII,MACKKL,Fontenla+Avrett+Loeser1993}.  
However, the chromosphere is actually a very dynamic region.  
In this paper we investigate the
hydrogen ionization structure of a dynamic chromosphere.  

It has long been known that hydrogen ionization and excitation in the 
solar atmosphere is not in equilibrium for the local temperature and 
density \citep{Thomas48}.  Usually, hydrogen ionization is calculated 
from the condition of ``statistical equilibrium'', that is, the 
equality of the ionization and recombination rates for the local 
temperature, density and radiation.  Statistical equilibrium assumes 
infinitely fast rates and an instantaneous adjustment to the local 
thermodynamic and radiation state.  However,
if the local state changes in time, or if there is a flow through 
a inhomogeneous region, and if the time scale to reach ionization or 
excitation equilibrium is longer than the dynamic times scale, then it 
is necessary to solve the population rate equations,
\UQ
{{d n_{i}}\over{d t}} = {\rm gains - losses} \ ,
\UN
for each species $i$ \citep{Joselyn+79}.  This latter is the case in the 
solar atmosphere heated by any intermittent process (such as shocks or
nanoflares)
\citep{Elzner75,Klein+Stein+Kalkofen76,Kneer+Nakagawa76,Poletto79,Kneer80}.  
The ionization and excitation state then depends 
on its history as well as the instantaneous temperature, density and 
radiation.  However, the importance of slow chemical reaction rates in 
the quiet solar atmosphere has generally been neglected since these 
early papers were written.  The active solar atmosphere (e.g. flares, 
prominence formation and coronal mass ejections) has long been recognized to 
require a time dependent analysis
\citep{McClymont+Canfield83,Doschek84,Fisher+Canfield+McClymont85,Heinzel91,%
Abbett+Hawley99,Sarro+99,Ciaravella+01,Ding+01,Lanza+01}.
The same issue of ionization and 
other chemical reaction rates being slow compared to the rate of
dynamical changes, and thus requiring the solution of the species rate 
equations rather than statistical equilibrium, arises in many other 
areas of astrophysics where conditions are changing in time.  Examples 
are: the interstellar medium \citep{Lyu+Bruhweiler96,Joulian+98},
HII regions \citep{Rodriguez-Gaspar+Tenoria-Tagle98,Richling+Yorke00}, 
planetary nebula 
\citep{Schmidt-Voight+Koppen87,Frank+Mellema94,Marten+Szczerba97}, 
novae \citep{Hauschildt+92,Beck+95}, supernova \citep{Kozma+Fransson98}, 
and the reionization of the intergalactic medium and the Ly$_{\alpha}$
forest 
\citep{Ikeuchi+Ostriker86,Shapiro+Kang87,Shapiro+94,Ferrara+Giallango96,%
Giroux+Shapiro96,Zhang+97}.

The solar chromosphere is not well represented by a static mean structure.
We show that because the chromosphere is dynamic the average 
hydrogen ionization is
significantly different from the ionization state of the mean atmosphere.
There are two reasons for this.  First, the mean of any quantity that
depends non-linearly on the atmospheric properties is not the same as
that quantity determined from the mean atmosphere.  Second, the finite
rates of ionization and recombination reduce the response of the
ionization to dynamic variations in the atmospheric state.  
Hydrogen does not have time to reach its equilibrium ionization.
The result
is a hydrogen ionization fraction that is higher than obtained for the
mean of the dynamic atmosphere.  This conclusion is
independent of just what that mean state is.  Thus analyzing
observations on the basis of a static model chromosphere leads to very
different conclusions from an analysis based on a dynamic model
chromosphere.  This significantly alters the interpretation of observations.

In this paper, we first describe (section 2) the numerical
simulations used to study the dynamic hydrogen ionization.  We then
show that the dominant hydrogen ionization process
is photoionization from the second level and we investigate the
processes that populate the second level (section 3).
There is a very rapid equilibration of the 2$^{nd}$ and all higher 
levels with the continuum, with a slow collisional leakage of electrons 
from the ground state to the 2$^{nd}$ level, or visa versa, depending 
on whether hydrogen is ionizing or recombining.  
Next we calculate 
the time scale for ionization/recombination and its relation to the
eigenvalues of the rate matrix (section 4).  The ionization
and recombination rates are slow and not in equilibrium, so that the
level populations change in time.  
We demonstrate that there is only one relaxation time scale to approach 
equilibrium, not separate ionization and recombination times scales.
We show (section 5) that
because of the slow ionization and recombination rates, the dynamic
ionization fluctuates much less than the statistical equilibrium
ionization.  As a result, line and continuum intensity variations do not
mimic the underlying dynamics.  Further, the mean ionization fraction 
tends to represent the maximum statistical equilibrium ionization, and is 
much higher than is obtained from the statistical equilibrium of the mean
atmosphere.  We conclude (section 6) with a reiteration of why the
chromosphere must be dynamic based on fundamental physical principles
and observations and a statement of which results are robust and why.

\section{Numerical Simulations}\label{sec:numerics}

To properly model the dynamic solar chromosphere one has to perform
radiation-hydrodynamic simulations taking into account the
non-local, non-linear rate equations for all important species.
Such a self-consistent radiation-hydrodynamic modeling of the solar
chromosphere has been performed by Carlsson \& Stein (1992, 1994, 1995, 1997a, 1997b)
\nocite{Carlsson+Stein1992}
\nocite{Carlsson+Stein1994}
\nocite{Carlsson+Stein1995}
\nocite{Carlsson+Stein1997}
\nocite{Carlsson+Stein1997b}
and we summarize the methodology here.

We solve the one-dimensional equations of mass, momentum, energy and
charge conservation together with the non-LTE radiative transfer and
population rate equations, implicitly on an adaptive mesh.  Advection
is treated using Van Leer's (1977) second \nocite{vanLeer1977} order
upwind scheme to ensure stability and monotonicity in the presence of
shocks.  An adaptive mesh is used \citep{Dorfi+Drury} in order to
resolve the regions where the atomic level populations are changing
rapidly (such as in shocks).  The equations are solved simultaneously
and implicitly to ensure self-consistency and stability in the presence
of radiative energy transfer, stiff population rate equations, and to
have the time steps controlled by the rate of change of the variables
and not by the small Courant time for the smallest zones. A linearization
method is used to solve the radiative transfer \citep{Scharmer+Carlsson1985}
but with a penta-diagonal approximate lambda operator \citep{Rybicki+Hummer1991} instead of the global Scharmer operator.
The effects of
non-equilibrium ionization, excitation, and radiative energy exchange
from several atomic species (H, He and Ca) on fluid motions and the
effect of motion on the emitted radiation from these species are
calculated.  We model hydrogen and singly ionized calcium by 6 level
atoms and helium with a 9 level atom. For helium we collapse terms to
collective levels and include the $1s^2$, $2s$ and $2p$ terms in the
singlet system and the $2s$ and $2p$ terms in the triplet system of
neutral helium, and the $1s$, $2s$ and $2p$ terms of singly ionized
helium. In addition we include doubly ionized helium.
We include in detail all
transitions between these levels. 
For singly ionized calcium they are the H and K resonance
lines, the infrared triplet and the photoionization continua from the
five lowest levels. 
We use 31--101 frequency points in each line and 4--23 frequency
points in each continuum; a total of 1424 frequency points.
Continua from elements other than H, He and Ca 
are treated as background continua
in LTE, using the Uppsala atmospheres program \citep{Gustafsson1973}.

The upper boundary is a corona at 10$^6$K with a transmitting boundary
condition.  Incident radiation from the corona is included which causes
ionization in the helium continua in the upper chromosphere. The lower
boundary condition, at 500\,km below $\tau_{500}\!=\!1$, is also
transmitting.  Waves are driven through the atmosphere by a piston
located at the bottom of the computational domain.  
The piston velocity is chosen to reproduce a 3750 second sequence of
Doppler-shift observations in an Fe~I line at $\lambda$396.68\,nm in
the wing of the Ca~H-line (Lites, Rutten \& Kalkofen 1993)
\nocite{Lites+Rutten+Kalkofen1993a} 
This line is formed about 260\,km above $\tau_{500}\!=\!1$.

\begin{figure}
\plotone{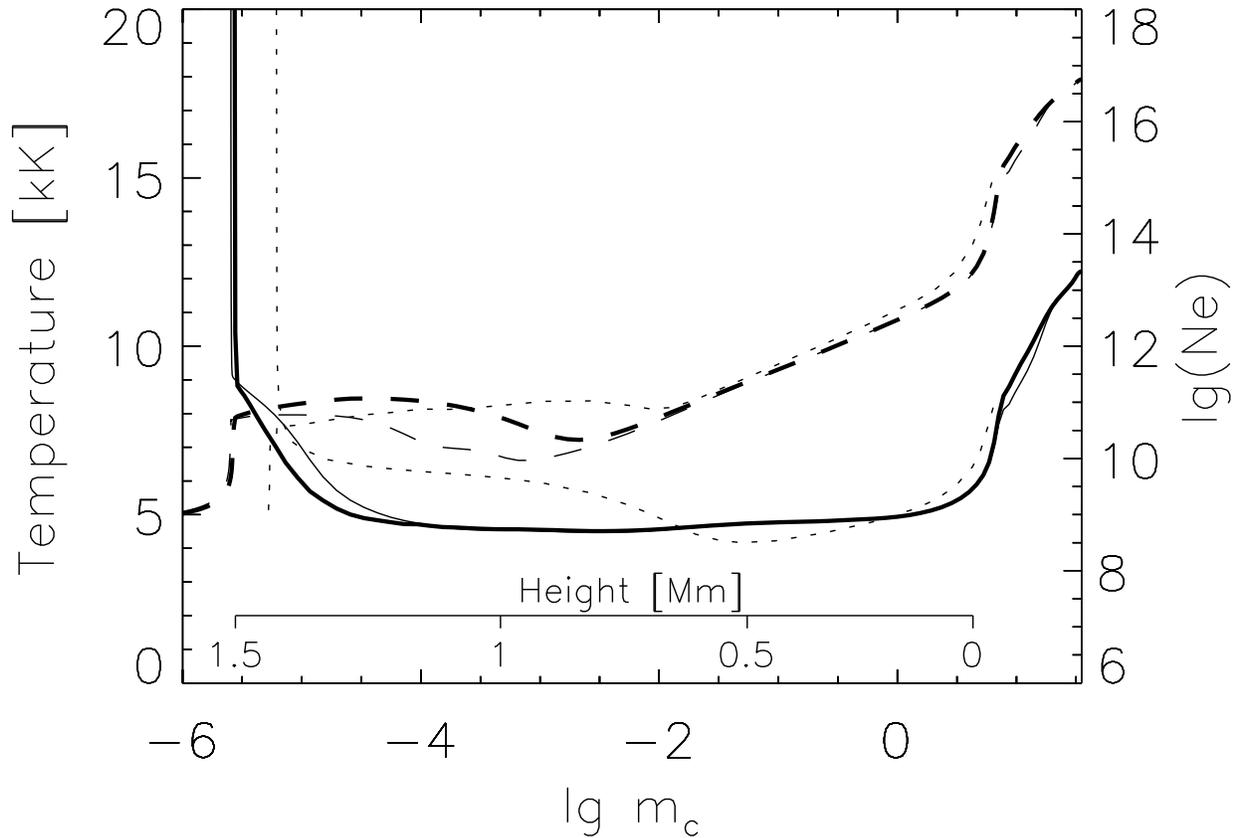}
\caption{
Initial and mean state of the atmosphere. Temperature (solid lines) and
electron density (dashed lines) as functions of column mass.
The initial state is shown with thin lines and the mean of the second half
of the simulation is shown with thick lines. The height scale is that of
the initial atmosphere. The mean temperature is close
to the initial radiative equilibrium state while the mean electron density
is up to a factor six larger than the initial value. For comparison, the
temperature and electron density as functions of column mass
of the VALIIIC model are shown with dotted lines.
\label{fig1}}
\end{figure}

The initial atmosphere (Fig.~\ref{fig1}) is in radiative
equilibrium above the convection zone (for the processes we consider)
without line blanketing and extends 500\,km into the convection zone,
with a time constant divergence of the convective energy flux (on a
column mass scale) calculated with the Uppsala code without line
blanketing.  Note that the transition region in the initial
atmosphere occurs at a lower height than in standard models.  This is
because the lower temperature means a smaller pressure scale height
and a less extended atmosphere. The atmospheric dynamics
moves it outwards on average. The transition region occurs at a
smaller pressure than in the VALIIIC model. This is set by the
amount of conductive flux at the upper boundary and is a free
parameter in the model (follows from the location of the upper 
boundary at the fixed temperature of 10$^6$K).
The mean structure of the dynamic atmosphere (Fig.~\ref{fig1})
has a low (5000 K) temperature throughout most of the chromosphere, with
a temperature rise in the upper chromosphere produced by absorption of
coronal radiation in the helium continua.

The difference between the calculations reported here and the ones in
the references cited above is that we now include helium, a corona
and transition region, the incident radiation from the corona, extend
the calculations deeper and have a transmitting lower boundary
condition.  This makes it possible to discuss the upper chromosphere,
in particular the ionization of hydrogen, in some more detail than
before.

The validity of these simulations can be checked by comparing their
predictions with observations.  There is both agreement with some 
observations and disagreement with other observations.
CO molecular line observations indicate that there is no
temperature rise in the low chromosphere, consistent with the
simulation.  
The simulations reproduce many details in observations of the calcium
H-line \citep{Carlsson+Stein1997}.  The Ca grains are due to waves
that steepen into shocks around a height of 1 Mm.  
However, the cores of the simulated H and K
lines are darker and the bright points are brighter than is observed.
The Lyman continuum intensity variation is also larger in the
simulations than what is observed. 
There is a possible disagreement with SUMER observations
that the cores of NI, OI, CI and CII lines, formed in the
mid and upper chromosphere, show emission 
everywhere.  Although some of these line cores are formed below the height
where coronal radiative heating raises the mean chromospheric
temperature, there is some contribution from that region of
enhanced temperature. Preliminary calculations (that are non-trivial
because effects of slow ionization/recombination have to be included, as
is shown in this paper) show that this contribution actually cause the
lines to be in emission all the time even though the average line emission is
weaker than is observed.

The improved physics in the simulations reported on here compared with
the previous work (the inclusion of a corona and the absorption of
coronal radiation in the helium continua) has improved the agreement;
the line cores in the calcium H and K lines are not as dark as 
in the previous simulations, the rms variation in the Lyman continuum
radiation temperature has changed from 256~K 
(Carlsson \& Stein 1994, 1997b) to 97~K (compared to an observed value
of about 40~K) and UV lines from neutral elements now seem to be in
emission everywhere in the simulations.

The existing disagreements indicate that there is still physics
missing from the simulations. 
For instance, we do not
include line blanketing (especially the numerous iron lines), the CO 
molecule and the singly ionized magnesium atom (producing the Mg h and k
lines).  We use a crude approximation to partial
redistribution of radiation in the Lyman lines and no partial
redistribution at all in the calcium H and K lines.
We do not include frequencies above 20 mHz in
the driving piston. 
We do not include the effects of magnetic fields.

%
Assuming complete redistribution in the calcium lines probably leads
to an overestimate of the cooling in these lines by up to a factor of
two \citep{Uitenbroek2002}, partially compensating the omission of the
magnesium h and k lines. 
High frequency waves  may lead to heating 
of the mid chromosphere through dissipation in shocks and the neglect
of high frequencies in the driving piston probably leads to a lower
mean temperature. 

Magnetic fields structure the chromosphere and corona 
\citep{Aschwanden+01,Berger+99,Schrijver+99,Moses+94}.  
They influence both energy transport (wave modes and
quasi-static driving) and heating (nanoflares, resonant wave
absorption).  In the photosphere magnetic fields are concentrated into
isolated flux tubes or loops, except in very high magnetic flux
regions, leaving most of the photosphere nearly field free.  The
upper convection zone is the site of acoustic wave generation
\citep{Stein+Nordlund01,Skartlien+00,Goldreich+Murray+Kumar94}, while overshooting
convective flows in the photosphere generate gravity waves.  MHD tube
waves are driven by convective motions acting on the localized magnetic 
fields \citep{Musielak+Ulmschneider01}.
The photospheric tubes and loops spread out with increasing height and
decreasing gas pressure.  When acoustic waves propagate into the the
region where the magnetic pressure equals the gas pressure 
($\beta = 1$) and the magnetic field lines are curved, there is 
significant reflection and mode conversion 
into magneto-hydrodynamic waves propagating at approximately the Alfven speed
\citep{Rosenthal+etal2002}.  
These latter are less compressive and
have a smaller amplitude for a given flux than the acoustic waves, 
because of their faster propagation speed, and hence are less visible.
Where there is magnetic field extra heating occurs 
--- continuum intensities and line emissions are
substantially higher in magnetic network regions than in the
internetwork.  It is likely that the same processes will
contribute to the heating in internetwork regions in the
mid-chromosphere where the magnetic field spreads to form a "magnetic
canopy".


In this paper we focus on the
processes controlling the ionization of hydrogen in a dynamic atmosphere.
We show that when chemical rates are slow, populations do not have
time to reach their equilibrium values.  Indeed, some of the
disagreement with observations may be due to this non-equilibrium.
The solar chromosphere will experience some additional heating
besides that included in our model.  However, as we show later, even in
such hotter models the ionization/recombination rates are slow and the
results presented here will be qualitatively correct.

\section{Ionization and Excitation Processes}\label{sec:processes}

In the upper chromosphere and lower transition region the density is much too
low to ensure LTE and the full rate equations have to be solved to
calculate the hydrogen ionization. Furthermore, inspection of the
rates involved shows that typical ionization/recombination time scales
are much longer than dynamical time scales so that the ionization
balance can be expected to be out of statistical equilibrium. It is
thus necessary to take into account the advection and time-derivative
terms in the equations.
In this section we analyze the rate equations to find what processes
dominate the hydrogen ionization balance. In the next section we
analyze the time scales involved.

\begin{figure}
\plotone{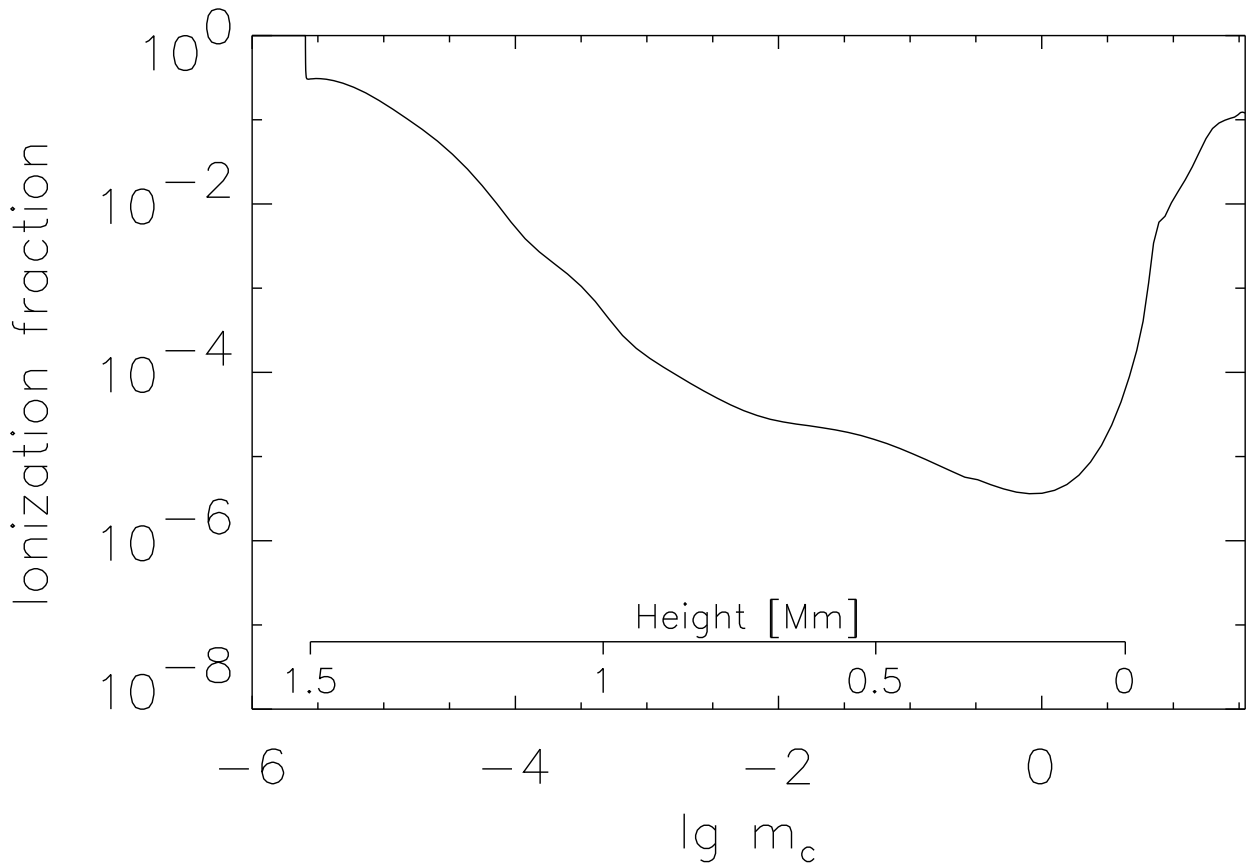}
\caption{
 Hydrogen ionization as a function of column mass in the initial
 radiative equilibrium atmosphere.The ionization fraction rises from
 about $10^{-5}$ at the classical temperature minimum at
 lg($m_c$)=$-1$ (height of 0.5 Mm) to about 30\% at the base of the
 transition region 1 Mm further up.
\label{fig2}}
\end{figure}

Figure~\ref{fig2} shows the hydrogen ionization fraction as a
function of column mass ($m_c$) in the initial radiative equilibrium
atmosphere. The ionization fraction rises from about $10^{-5}$ at the
classical temperature minimum at lg($m_c$)=$-1$ (height of 0.5 Mm) 
to about 30\% at the base of the transition region 1 Mm further up.

\begin{figure}
\plotone{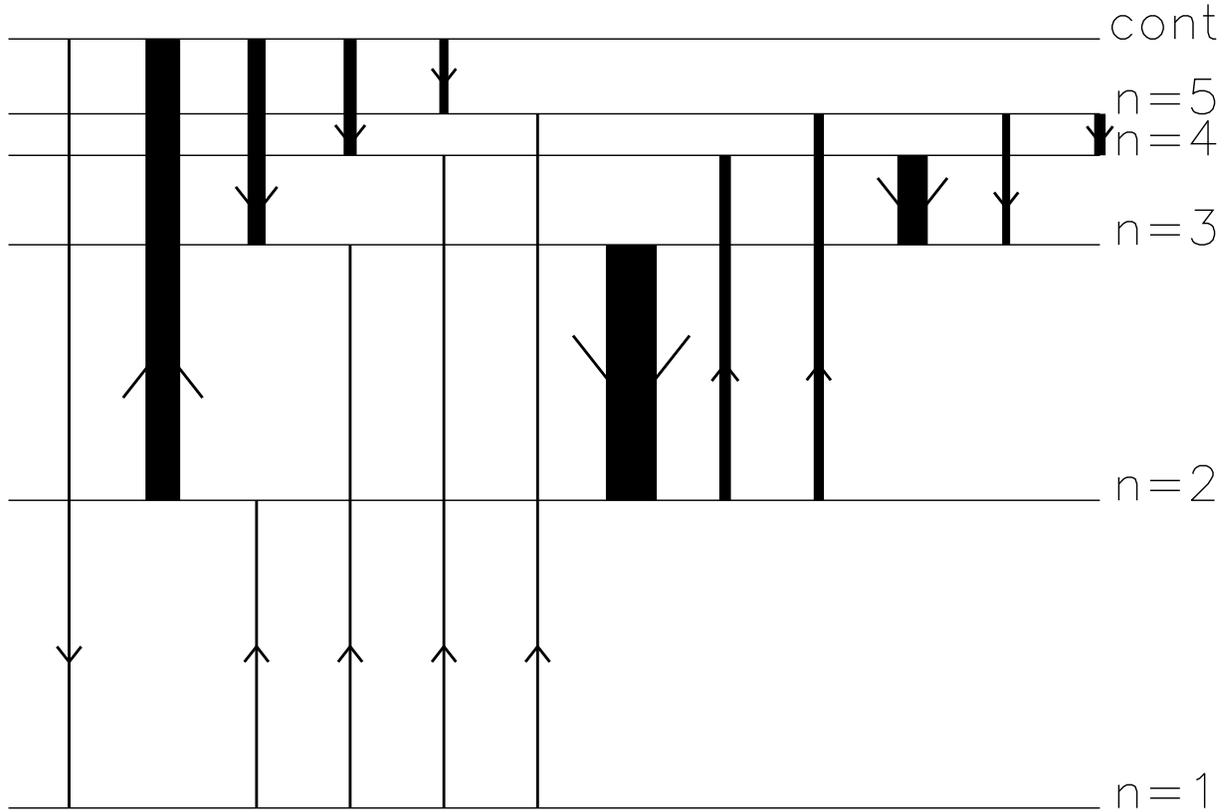}
\caption{
  Net rates between the hydrogen levels in the chromosphere
  (at lg($m_c$)=$-4$ which corresponds to a height of 1.2 Mm in the initial 
  atmosphere, but the picture is almost identical in the whole chromosphere 
  from lg($m_c$)=$-1$ up to the base of the transition region at 
  lg($m_c$)=$-5.6$).  The thickness
  is proportional to the net rate with arrows showing the direction of the
  net rate. The ionization is dominated by photoionization in the Balmer
  continuum balanced by photorecombination to higher levels and cascades
  through bound-bound transitions with $\Delta n$=1.
  \label{fig3}}
\end{figure}

The processes that are important for the ionization balance are not
the same in the transition region and in the
chromosphere. 
In the solar chromosphere the hydrogen ionization is dominated by
photoionization from the first excited state. 
The reason is that the much lower particle density of the first excited 
state ($\approx 3 \times 10^4 {\rm cm}^{-3}$) than in the ground state 
($\approx 10^{11} {\rm cm}^{-3}$) is more than compensated by the higher
mean intensity at the Balmer edge at 364 nm compared to that
of the Lyman edge at 91 nm. At a radiation temperature of 5000~K the
ratio is $3 \times 10^8$. In addition, the atmosphere is optically
thick to Lyman photons and optically thin to Balmer photons in all of
the chromosphere. The result is that photoionization by Lyman photons
plays an insignificant role in the hydrogen ionization throughout the
chromosphere.
Figure~\ref{fig3} shows the net rates between
all the hydrogen energy levels at lg($m_c$)=$-4$ which corresponds to
a height of 1.2 Mm in the initial atmosphere. This picture is almost 
identical in the whole
chromosphere from lg($m_c$)=$-1$ up to the base of the transition region
at lg($m_c$)=$-5.6$. Ionization is primarily through a net rate from
$n$=2 to the continuum (photoionization in the Balmer continuum) balanced
by recombination to the higher levels cascading down to the $n$=2 level
through bound-bound transitions with $\Delta n$=1.
The reservoir of electrons is the ground state.  There is transfer of
electrons between the ground state and first excited state by
collisions.  The rate of this transfer is very slow, because
chromospheric temperatures are of order 1 eV and the energy jump to the
first excited level is 10.2 eV.

\begin{figure}
\plotone{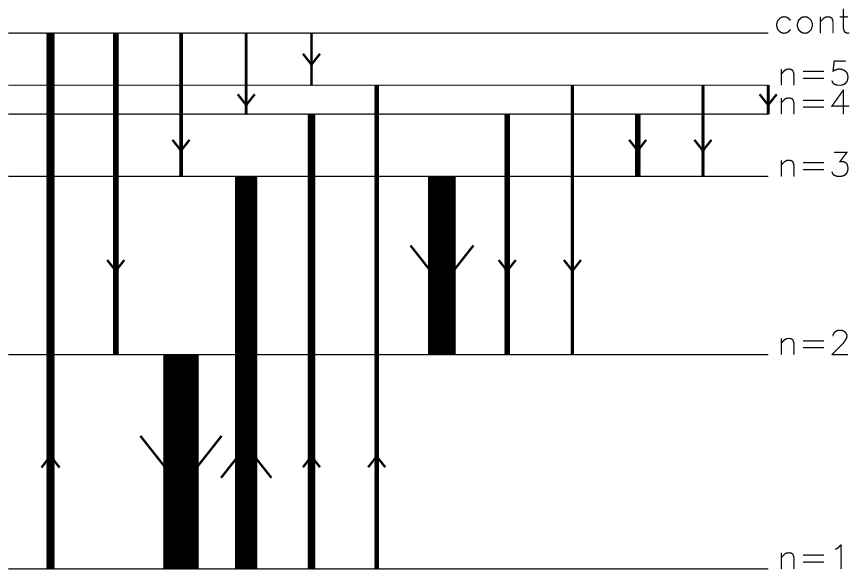}
\caption{
  Same as Fig.~\ref{fig3} but for a point
  in the transition region where T=20000K. The ionization is dominated
  by collisional ionization from the ground state balanced by 
  photorecombination to all levels.
  \label{fig4}}
\end{figure}

In the very thin zone (2~km) where the ionization
fraction goes from 30\% to fully ionized the situation is
different. Collisional ionization from the ground state dominates with net
photorecombination in all continua and again bound-bound transitions
with $\Delta n$=1 dominating the return channel to the ground state
(Fig.~\ref{fig4}).

Figures~\ref{fig3}-\ref{fig4} only show the net
rates and do not answer the question what is {\em driving} the system
and what rates are just adjusting to provide a closed loop. 
One way to investigate this aspect is to see how the
system responds to perturbations.  We have therefore solved the
equations of statistical equilibrium repeatedly perturbing the rates
one-by-one. 
Increasing the photoionization
cross-section in the Balmer continuum increases the ionization while
the opposite is true for the bound-free transitions from the higher
levels, consistent with the rate picture where the ionization is
through photoionization in the Balmer continuum and recombination to
the higher levels. Even though the H-$\alpha$ ($n$=3 $\rightarrow$ $n$=2)
transition has the largest net rate in the chromosphere changing this
rate has no effect on the ionization balance.  The same is true for all
other bound-bound transitions except for Ly${\alpha}$.
Increasing the Ly${\alpha}$ rate increases the ionization slightly just below
the transition region where absorption in Ly${\alpha}$ of photons from
the transition region and corona populates level 2.

\begin{figure}
\plotone{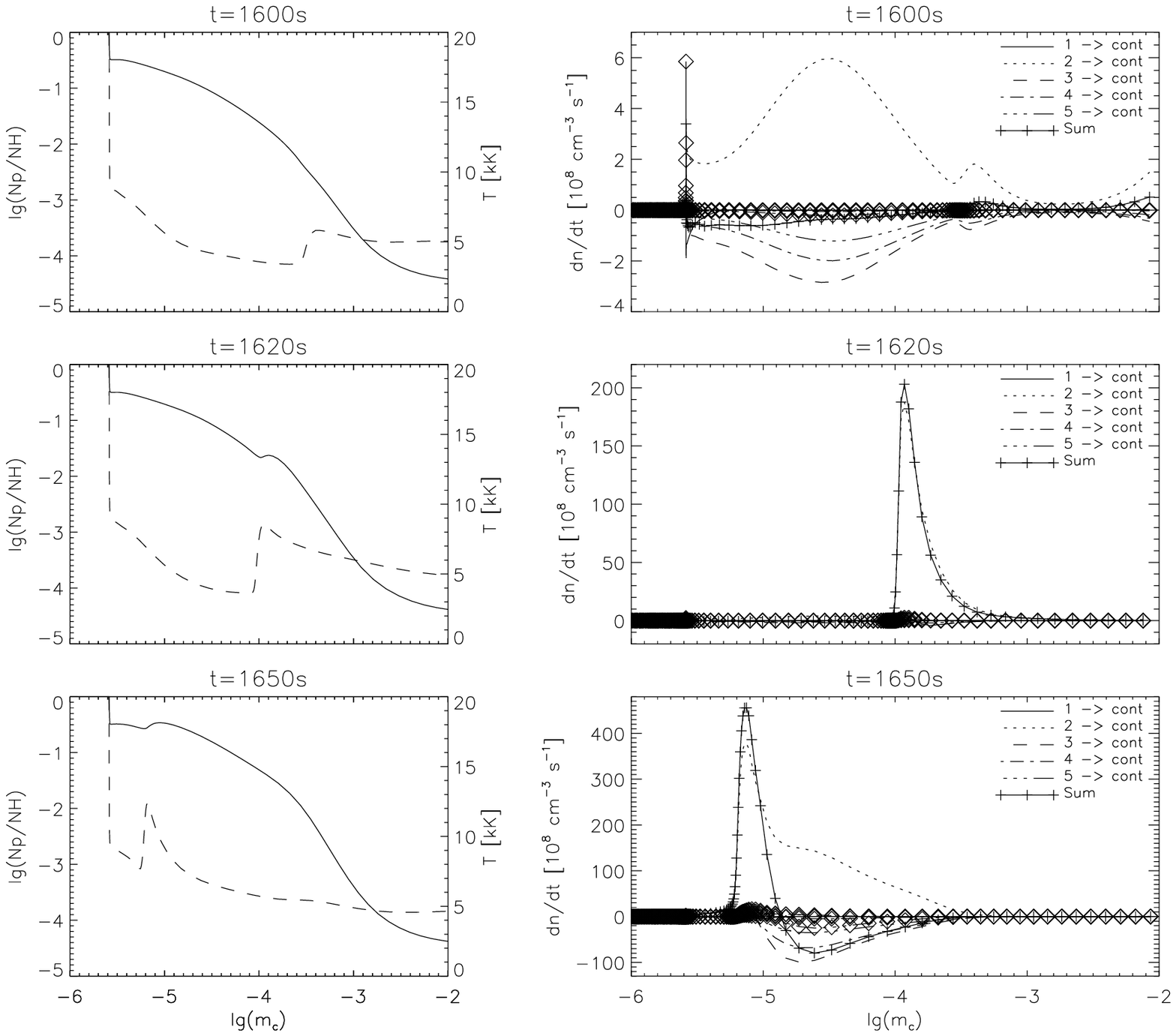}
\caption{
  The temperature (left column, dashed line, right axis) and
  hydrogen ionization fraction (left column, solid line, left axis)
  together with the rates (right column) as a function of column mass and
  time in a simulation of the solar chromosphere and lower transition
  region. Rates are given from the bound levels $n$=1-5 to the
  continuum with no symbols on radiative rates and diamonds added for
  collisional rates. Positive rates indicate ionization, negative
  rates recombination. Before the shock forms the rates are in
  equilibrium (plus signs close to zero) dominated by photoionization
  in the Balmer continuum and recombination to the higher levels (top
  right panel). As the shock progresses upwards the rates become
  larger and Balmer
  photoionization dominates without a balancing recombination to the
  highly excited levels.
  Direct ionization from the ground state is always small below the
  transition region.
\label{fig5}}
\end{figure}

Up to now we have studied only the static atmosphere.
Hydrogen ionization in a dynamic chromosphere is
illustrated in Fig.~\ref{fig5}. In the
left column the temperature is shown as a function of column mass as a
dashed line with the values shown on the right hand axes. The same
column shows the hydrogen ionization fraction on a logarithmic scale
as a solid line with values on the left axes. The right column shows
the net rates from the bound hydrogen levels $n$=1-5 to the continuum with
positive values indicating net ionization, negative values recombination.
Time progresses from the top panel downwards.

At a time 1600 seconds from the start of the simulation a
shock is starting to form around lg($m_c$)= $-3.5$ (height of 1 Mm). 
The rates are in statistical
equilibrium (the sum of the rates is close to zero) with
a balance between Balmer photoionization and recombination to the levels
$n$=2-5. 20 seconds later (mid panels) the shock has formed. Balmer
photoionization is greatly increased in the shock, because the population 
of the first excited state has greatly increased due to collisional
excitation from the ground state .  (The ratio of collisional
excitation to de-excitation rates depends exponentially on temperature,
$C_{12}/C_{21} = g_2/g_1 e^{- \Delta E / kT}$.)  The Ly${\alpha}$ transition
is found to be
in detailed balance and so just bounces electrons up and down.  The
photoionization rate per atom is constant because it depends on the
radiation field which is set deeper down in the atmosphere.
Balmer photoionization dominates the rates into the
continuum (dotted curve), but it {\em is not} balanced by recombination
(because of slow rates) resulting in a net increase of protons (sum of
rates positive and almost equal to the Balmer photoionization). As the
shock progresses farther upwards, recombination dominates over
photoionization behind the shock. 
The zone where hydrogen goes from 1\% to 40\% ionization is about 600
km thick.

At the top of the chromosphere hydrogen is between 30 and 40\%
ionized.  The further ionization up to 100\% ionization takes place in
the zone where the temperature increases from 10000~K to
25000~K.  The ionization rates are here dominated by collisional
ionization directly from the ground-state while the Lyman continuum
actually provides some {\em net recombination}. The thickness of this
zone depends on the temperature gradient and is thus very model
dependent --- in the plane parallel calculations this zone is very thin,
on the order of 2~km. Also in this zone we see rates that are out of
statistical equilibrium with net ionization when shocks pass followed by 
net recombination.

\section{Time Scales}\label{sec:timescales}

The crucial result of our investigation is that the time scale for
changes in the ionization of hydrogen in the chromosphere is long,
because the rates for ionization and recombination under conditions
typical of the solar chromosphere are small.  Consequently the hydrogen
populations do not adjust to the local conditions.

The relation of time scales and transition rates can be understood most
readily for the case of a 2-level atom.  We summarize the well known
results here.  The equations for the populations of levels 1 and 2 are
\EQA
{{D n_1}\over{Dt}} & = & n_2 P_{21} - n_1 P_{12} \\
n_1 + n_2 & = & n_H  \ ,
\ENA
where $n_i$ is the population of level $i$ and $P_{ij}$ is the
transition rate per atom (radiative $R_{ij}$ plus collisional $C_{ij}$)
from level $i$ to level $j$, and $n_H$ is the total hydrogen number
density.  The solution for the population of level
$1$ is
\EQ
n_1(t) = n_1(\infty) + \left(n_1(0) - n_1(\infty)\right)
e^{-t(P_{21}+P_{12})} \ ,
\EN
where $n_1(\infty)$ is the equilibrium population achieved at infinite
time,
\EQ
n_1(\infty) = {{n_H P_{21}}\over{P_{21}+P_{12}}} \ .
\EN
Thus there is only one time scale for for the approach to ionization equilibrium 
\EQ
\tau_{relax} = 1/\left(P_{21}+P_{12}\right) \ .
\EN
It is wrong to talk about separate time scales, one for ionization and 
one for recombination, even though there are distinct ionization and
recombination rates.  Also, this result of an exponential approach to a
final equilibrium state with a given time scale assumes that the
ionization and recombination rates per atom are constant in time.  In the real
case where the rates per atom are themselves evolving (because of
changes in the radiation field and electron density) the concept of a
time scale is not well defined, although it is still useful.

When the rate of upward radiative transitions balances the rate of 
downward radiative transitions, $n_1 R_{12} = n_2 R_{21}$, the equations
can be simplified by assuming that this holds exactly, which is called
{\it detailed balance}.
In this case, $n_i P_{ij} \rightarrow n_i C_{ij}$ and the solution for the 
population of level one is
\EQ
n_1(t) = n_1(\infty) + \left(n_1(0) - n_1(\infty)\right)
e^{-t(C_{21}+C_{12})} \ ,
\EN
where now the final equilibrium population is
\EQ
n_1(\infty) = {{n_H C_{21}}\over{C_{21}+C_{12}}} \ ,
\EN
and the relaxation time scale is
\EQ
\tau_{relax} = 1/\left(C_{21}+C_{12}\right) \ .
\EN
This assumption of detail balance thus implies that the radiative
rates per atom change with time as the populations relax to their
equilibrium state, so that
\EQ
{{R_{12}}\over{R_{21}}} = {{n_2}\over{n_1}} = {{n_2(\infty) - 
\left(n_1(0) - n_1(\infty)\right)e^{-t(C_{21}+C_{12})}}\over
{{n_1(\infty) + \left(n_1(0) - n_1(\infty)\right)e^{-t(C_{21}+C_{12})}}}}
\ .
\EN

Another possible simplification is to express the radiative rates in
terms of the {\it net radiative bracket} (NRB),
\EQ
n_2 R_{21} - n_1 R_{12} = n_2 R_{21} \left(1 -
{{n_1}\over{n_2}}{{R_{12}\over{R_{21}}}}\right) 
= n_2 R_{21}\cdot{NRB}
\ .
\EN
In this case the equation for the level 1 population becomes
\EQ
{{D n_1}\over{Dt}}  =  n_2 \left(C_{21}+R_{21}\cdot{NRB}\right) - n_1 C_{12}
\ .
\EN
If it is assumed that both the rates per atom and the NRB are constant, then 
the solution for the population is
\EQ
n_1(t) = n_1(\infty) + \left(n_1(0) - n_1(\infty)\right)
e^{-t(C_{21}+C_{12}+R_{21}\cdot{NRB})} \ ,
\EN
where now the final equilibrium population is
\EQ
n_1(\infty) = n_H {{C_{21}+R_{21}\cdot NRB}\over{C_{21}+C_{12}+R_{21}\cdot NRB}} \ ,
\EN
and the time scale to approach equilibrium is
\EQ
\tau_{relax} = 1/\left(C_{21}+C_{12}+R_{21}\cdot{NRB}\right) \ .
\EN
If the net radiative transition rate is small compared to the upward
and downward rates individually, then the NRB will be very small, so the
radiative rates will make only a small change in the relaxation time
scale given by the collisional rates.  The radiative transitions can
either increase (if NRB $<$ 0) or decrease (if NRB $>$ 0) the relaxation
time scale.

We have calculated the relaxation time scale (at each height and each
time step) from our numerical simulation. We proceeded in the
following way: The value of the hydrodynamic variables were taken from
a given time step and kept constant in time for the relaxation time
scale calculation. The population densities consistent with this state
of the atmosphere were calculated by solving the equations of
statistical equilibrium. This defined the initial population density
in the ionized state, $n_p(0)$. This atmosphere was then perturbed by
increasing the temperature by 1\% throughout. The populations
consistent with this perturbed state was calculated from the equations
of statistical equilibrium giving the asymptotic solution,
$n_p(\infty)$. The time evolution from the initial state of the number
of protons at a given height was also calculated using the full rate
equations, defining $n_p(t)$. The numerical solution was cast in a
2-level form (see above) and the relaxation time scale was calculated
from a least-squares linear fit to
\EQ
  \ln\left[{{n_p(t)-n_p(\infty)}\over{n_p(0)-n_p(\infty)}}\right]=
  -{t\over \tau_{relax}}
\ .
\EN

\begin{figure}
\plotone{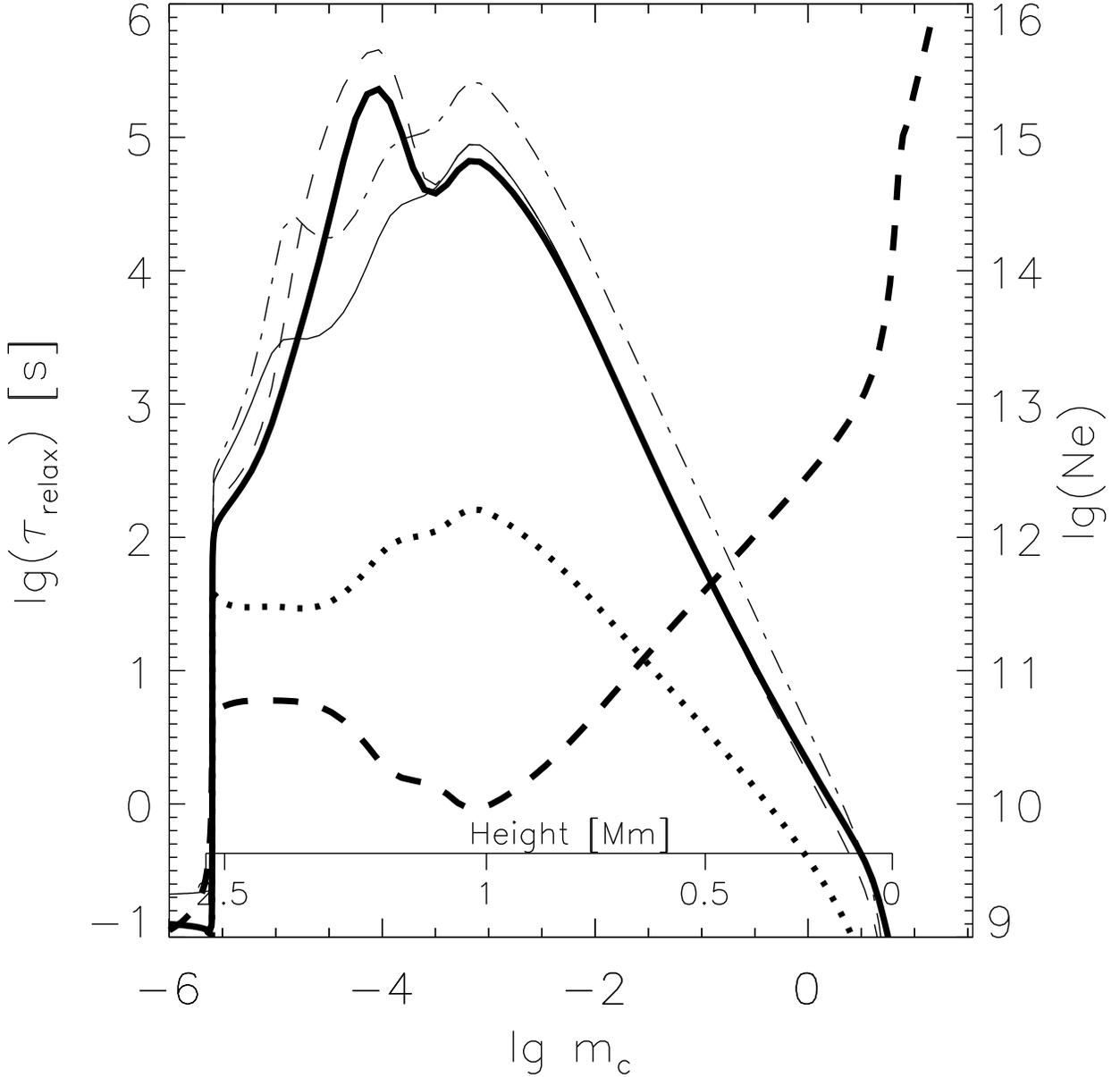}
\caption{Relaxation time scale as function
of column mass. The numerically determined relaxation time scale is given
as the thick solid line. Time scales determined from eigenvalues of the
rate matrix are also shown for several cases: full rate matrix (dotted), collisions
only (dot-dashed), Lyman transitions in detailed balance (thin solid) and
Ly$\alpha$ treated with a constant net radiative bracket (dashed). The
electron density is given as the thick dashed curve.
The ionization/recombination time scale becomes very long in the
chromosphere.  Eigenvalues calculated from a rate matrix with all the
radiative and collisional rates give a time scale several orders of
magnitude too short.  Eigenvalues calculated from a rate matrix with
all Lyman radiative transitions in detailed balance except for
Ly$\alpha$, which is included using its net radiative bracket, give a time
scale that closely matches the numerical result.
\label{fig6}}
\end{figure}

The relaxation time scale for hydrogen ionization/recombination, as
found from the numerical simulation, from the photosphere to the
transition region, is shown in Fig.~\ref{fig6} (thick
solid line).  The time
scale increases outward from the $\sim 1$ sec. in the photosphere to
$\sim 10^5$ sec in the mid chromosphere and then decreases to $\sim
10^2$ sec. at the base of the transition region.  

The eigenvalues  and eigenvectors of the rate matrix, 
$P_{ij}$ with $P_{ii} = \sum_j P_{ij}$ (where $P_{ij}$ is the transition
rate per atom from level $i$ to level $j$), help to clarify 
the processes controlling the time scale.  
The relaxation process can be represented by the equation
\EQ\label{eqn:relax}
\mathbf{n} = \sum_{i=0,5} c_i \mathbf{v_i}
e^{\lambda_i t} \ ,
\EN
where $\mathbf{n}$ is the vector of the level populations,
\EQ
\mathbf{n} = \left(\begin{array}{c}
             n_1\\n_2\\n_3\\n_4\\n_5\\n_6
             \end{array} \right) \ .
\EN
Here level 1 is the ground state and level 6 is the continuum.
$\mathbf{v_i}$ is the eigenvector corresponding to the $i^{th}$
eigenvalue $\lambda_i$.  The coefficients $c_i$ depend on the initial
conditions.  
There is a zero eigenvalue whose eigenvector is the equilibrium state.
The other eigenvalues are all negative, since the populations are
relaxing toward their equilibrium value.  

The numerically determined relaxation time scale is compared with the
time scale from the eigenvalue calculation, which is
the inverse of the smallest (in absolute value) non-zero eigenvalue, 
in Fig.~\ref{fig6}.  We have calculated the eigenvalues
using several different assumptions about the rate matrix.
Note, first of all, that when all the processes are included in the rate
matrix the time scale obtained from the eigenvalues (dotted line) is 
orders of magnitude smaller than found in the numerical
solution (thick solid line).  The time scale obtained from the rate
matrix with only collisional rates (dot dash line) has the same general 
pattern as the
numerical time scale, but is generally slightly larger.  When radiative
rates are added, but with all the
Lyman transitions assumed to be in detailed balance, the time scale from
the eigenvalues of the rate matrix (thin solid line) 
reproduce the numerical time scale up to the peak in the
mid-chromosphere.  This indicates that in this region the large Lyman radiative
rates are in fact changing with time so as to maintain nearly detailed
balance as the populations change.  The general behavior of the 
relaxation time is thus controlled by the collisional processes.  The
time scale increases from the photosphere to the mid chromosphere,
where the electron density has a minimum approximately as the inverse of
the collisional recombination rate ($\propto n_e^{-2}$). 
From the mid-chromosphere to the transition region the time
scale decreases, first because the electron density (thick dashed line) 
increases as the
increasing ionization fraction of hydrogen more than offsets the
outwardly decreasing overall density, and second because the
temperature starts to increase due to absorption of coronal photons.
The upward collisional rates are exponentially sensitive to the
temperature.  Radiative transitions generally reduce the time scale by
a factor of about 4 from the all collision case.  The wiggles in the
relaxation time above its peak in the
mid-chromosphere are nearly reproduced if the net Ly${\alpha}$ 
radiative transition is included as a constant {\it net radiative
bracket} (dashed line).  Where Ly${\alpha}$ photons begin to leak through the
atmosphere they increase the relaxation time because in that region
they produce a net increase in the population of the second level while
collisions produce a net depopulation of the second level.  Hence, the
Ly${\alpha}$ {\it net radiative bracket} is negative and as seen in
the two level case this increases the relaxation time.

The role of the various transitions in the equilibration process can be
determined from the eigenvectors and eigenvalues appearing in
eqn.~\ref{eqn:relax}.
We study the level in the atmosphere at column mass density
$m_c = 10^{-4}$ in the initial atmosphere, and treat the Lyman
transitions in detailed balance except for Ly${\alpha}$ which is
treated via its {\it net radiative bracket}. 
The slowest process, with eigenvalue $-2.2\times10^{-6} \, {\rm s}^{-1}$, 
has the eigenvector 
\EQ
\mathbf{v_1} = \left(\begin{array}{c}
             -0.7\\
		 2\times10^{-7}\\
		 1\times10^{-9}\\
		 4\times10^{-10}\\
		 3\times10^{-10}\\
		 0.7
             \end{array} \right) \ .
\EN
which thus represents 
the relaxation between the ground state and the continuum.  
The next slowest process, which is 10 orders of magnitude faster, 
with eigenvalue $-2\times10^4 \, {\rm s}^{-1}$, has the eigenvector  
\EQ
\mathbf{v_2} = \left(\begin{array}{c}
             -2\times10^{-4}\\
		 0.7\\
		 4\times10^{-3}\\
		 1\times10^{-3}\\
		 9\times10^{-4}\\
		 -0.7
             \end{array} \right) \ .
\EN
and is relaxation between the second level and the continuum.
The fastest relaxation processes, all with eigenvalues 
of order $-1\times10^7 \, {\rm s}^{-1}$, are between level 2 and levels 3, 
4 and 5.  
In the chromosphere, unlike the transition region, ground state
relaxation does not take place via direct level 1 $\leftrightarrow$ 
continuum transitions, but rather through a level 1 $\leftrightarrow$ 
2 transition followed by level 2 $\leftrightarrow$ higher levels and 
continuum transitions (Fig. \ref{fig3}).
Thus the process of hydrogen ionization in the chromosphere
can be thought of as a very rapid (small fraction of a second) 
equilibration of the 2$^{nd}$ and all higher levels with the continuum,
together with a slow leakage of electrons from the ground state to the
2$^{nd}$ level, or visa versa, depending on whether hydrogen is ionizing
or recombining.

\section{Consequences of Slow Rates}\label{sec:consequences}

\begin{figure}
\plotone{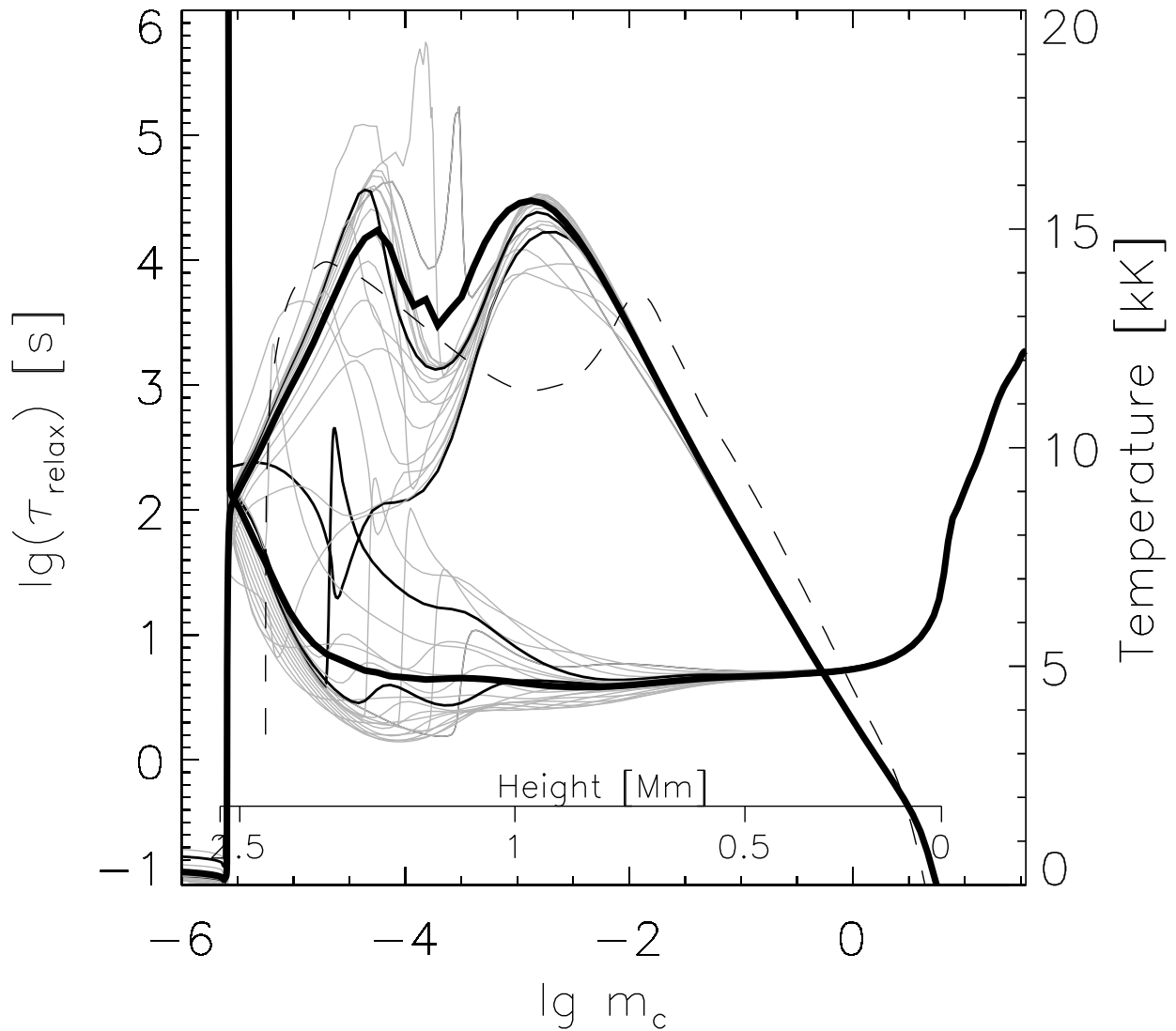}
\caption{Relaxation time scale 
(upper curves, scale to the left) and temperature (lower curves,
scale to the right) as function of
column mass and time in the dynamic simulation. The height scale of
the initial state is also given. The mean relaxation time scale and the
mean temperature over the second half of the simulation are given with
thick solid lines. Two particular instances are given in black with a
number of other instances in grey.
The time scale for ionization/recombination is much reduced in
shocks where hydrogen is ionizing and also reduced, but to a lesser
extent, in shock wakes where hydrogen is recombining. 
The relaxation 
time scale calculated from the VALIIIC semi-empirical model is given
as function of column mass for comparison (dashed line). 
\label{fig7}}
\end{figure}

The ionization structure of a dynamic atmosphere is very different from
that of a static atmosphere.  The ionization/relaxation time scale
decreases dramatically (to about 10 - $10^3$ sec.) in chromospheric
shocks, where the temperature increases significantly and to a lesser
extent in the elevated temperature tail of the shocks 
(Fig.~\ref{fig7}).  As a shock propagates upward it
strengthens and the time scale becomes shorter.  However, the time scale
is still too long for the ionization to reach its equilibrium value in
the shock and the peak ionization occurs behind the shock front
(Fig.~\ref{fig7}).  The slow ionization results in more of
the shock energy going into raising the temperature rather than ionizing
the gas \citep{Carlsson+Stein1992}.

\begin{figure}
\plotone{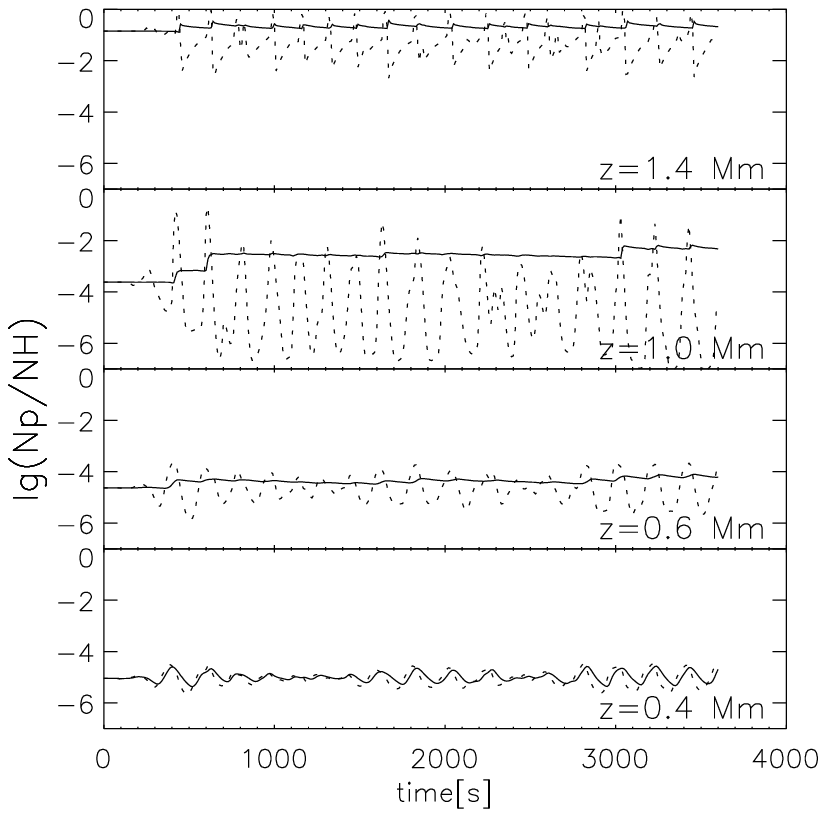}
\caption{Ionization fraction as function of time at four
   different Lagrangian locations (labeled with their initial height).
   The solid line shows the actual ionization fraction in the simulation,
   the dotted line shows the ionization fraction calculated from statistical
   equilibrium at the same values of the hydrodynamic variables.
   In the chromosphere, the ionization tends toward the equilibrium
   value appropriate for the shock peaks and its variation is small.
\label{fig8}}
\end{figure}

The primary processes controlling the ionization in shocks is the same as
for the static atmosphere:  photoionization from the Balmer level and
photorecombination to the higher levels, with slow leakage between the
ground state and the 2$^{nd}$ level.  The high temperature (and
density) in the shocks increases the collisional and radiative leakage
rates between the ground state and the 2$^{nd}$ level which leads 
to a smaller relaxation time.  Balmer photoionization has the greatest
increase, both within the shock and in the post shock tail 
(Fig.~\ref{fig5}).
In the shock itself, the sum of all the ionization/recombination rates
is approximately equal to the Balmer photoionization rate alone.  That
is, almost
all the ionization is occurring as Balmer photoionization and there is
little balancing recombination.  Thus the
ionization is not in statistical equilibrium and there is net ionizing
in the shock.  In the mid-chromosphere there is net ionization even in
the post shock tail, though at a slower rate.  When the shocks reach the
upper chromosphere, increased recombination to the third and higher
levels leads to net recombination in the post shock tail, but at a much
slower rate than the initial ionization in the shock 
(Fig.~\ref{fig5}, t= 1650 s).  Because the time scale
is short in the shock where hydrogen is ionizing, due to the high
temperature, but longer in the post shock region, where the temperature
decreases and hydrogen is recombining, the steady state level of hydrogen
ionization tends toward the value approximating the peak ionization in
the shocks (Fig.~\ref{fig8}).

In the photosphere (e.g. initial height of 0.4~Mm), the dynamical
ionization fraction follows the statistical equilibrium one with a
delay of 20-40 seconds and the ionization fraction is about $10^{-5}$
(Fig.~\ref{fig8}).
By the low chromosphere (initial height of 0.6~Mm) the rates become too
slow to keep up with the dynamic variations of density and temperature
and the ionization fraction shows much smaller variations than in
equilibrium. The ionization fraction increases with time from $2 \times
10^{-5}$ to a steady state value of $10^{-4}$.  In the mid-chromosphere
(initial height 1.0~Mm) the rates are so slow that the ionization
variation does not follow the dynamics at all.  There is a slow, secular
increase in the steady state ionization fraction from $10^{-4}$ to
$10^{-2}$. At the same values of the hydrodynamic variables the
statistical equilibrium ionization varies by six orders of magnitude.
The behavior slightly higher (at 1.4~Mm) is similar.  The slow
ionization/recombination rates thus cause the ionization fraction to
vary much less than in equilibrium and the mean ionization fraction
represents the conditions at the peaks of the shocks and is
substantially higher than the mean equilibrium value.

The small amplitude of the electron density variations compared with
their equilibrium values and the fact that the mean electron density
samples the peaks of the shocks rather than the mean conditions have
several consequences for the proper interpretation of chromospheric
diagnostics.

The intensity of collisionally excited lines will vary much less
than the hydrodynamic variations would imply. An analysis based on
equilibrium values of the electron density would give too low amplitudes
for the atmospheric variations even if the effects of departures from
LTE are taken fully into account.

Classical static models of the solar chromosphere are based on
temporal and spatial averages of intensity --- either continuum
intensities shortward of the silicon edge at 152 nm (like in the
models by Avrett and co-workers, e.g.,\ \cite{VALIII}, \cite{MACKKL},
\cite{Fontenla+Avrett+Loeser1993}) or line profiles of resonance
lines from ionized calcium and magnesium. 
The mean is taken in the ultraviolet part of the spectrum where
the temperature dependence of the Planck function is more exponential
than linear. Although the source function is quite decoupled from the Planck
function and has a smaller amplitude than the local Planck function, 
the mean source function preferentially samples the high
temperatures in shocks. This was dramatically illustrated in
\cite{Carlsson+Stein1995} where the best semi-empirical fit to the
mean intensities in the simulation showed a chromospheric temperature
rise while the mean temperature showed no increase. 

At long wavelengths the Planck function has a linear dependency on
temperature and one avoids the effect of averaging a non-linear
function (at ultraviolet wavelengths) that gives an exaggerated
temperature increase with height.
The temporal average of the intensity at mm and sub-mm wavelengths should give
the mean temperature as function of height provided the formation is
in LTE.  
However, the electron density sampling of
shock conditions at long wavelengths has a similar effect as temperature
sampling of shocks at short wavelengths and an
analysis based on equilibrium values will give an exaggerated
temperature
increase with height. The current simulations do, in fact, match
observations made at long wavelengths remarkably well
\citep{Loukitcheva+Solanki+Carlsson2001}.
The conclusion from the simulations is that you can not construct a model of
the mean atmosphere from temporally averaged intensities.

The result that the hydrogen ionization/recombination is slow is
robust. As detailed in section 2 the present simulations do not
contain several important physical ingredients for the proper
modelling of the upper chromosphere. However, the shocks propagating
through the chromosphere span all reasonable chromospheric conditions
and the time scale for ionization/recombination is always longer than
the dynamical time scale. Even the VALIIIC semi-empirical model
atmosphere with a temperature rise already at 0.5 Mm height and a
rather high temperature throughout the chromosphere gives an
ionization/recombination time scale around 10$^3$ seconds
(Fig.\ref{fig7}).

\section{Conclusion: The Dynamic Chromosphere}\label{sec:conclusions}

The solar chromosphere is a dynamic region.  This is obvious
both from basic properties of the solar atmosphere and from
observations.  First, the solar photosphere is continually perturbed by
convection.  The large drop in density between the photosphere and the
base of the transition region (14 scale heights) means that any
disturbance in the photosphere that propagates upward will grow in
amplitude with height in order to conserve energy unless it is strongly
dissipated.  Thus the chromosphere will experience large amplitude
motions driven from the photosphere.  Second, chromospheric spectral
lines such as CaII H and K and Mg h and k and the ultraviolet continuum
show significant temporal and spatial variability on a wide range of
periods and spatial scales.

Observations of such dynamic regions can not be properly analyzed based
on static model atmospheres.  There are four robust results from the
analysis of this paper that exemplify this basic truth and are 
independent of the details of the
specific model:  
First, the hydrogen ionization and recombination rates under any
reasonable
chromospheric conditions are slow compared to the rate of dynamical
changes there.  As a result, the ionization/recombination time scale is
longer than the dynamical time scale in the chromosphere.  Second, the
fluctuations in the hydrogen ionization are smaller than the variation
of its statistical equilibrium value calculated from the instantaneous
conditions.  Because the ionization/recombination time scale is longer
than the dynamical time scale, there is never enough time for the
ionization state to reach its equilibrium value.  Third, the hydrogen
ionization is greater than the statistical equilibrium value for the
mean atmosphere.  Because the ionization and recombination rates
increase with increasing temperature and density, ionization in shocks
is more rapid than recombination behind them.  Therefore, the
ionization state tends to represent the higher temperature of the
shocks.  Finally, the mean value of a dynamic property is not the
same as that property evaluated for the mean atmosphere.  Where 
adjustments are not instantaneous, a property depends on the history of
the atmosphere as well as its instantaneous state.  In addition,
observed properties are in general non-linear functions of the state of
the atmosphere.  For both reasons $<f(x)> \ne f(<x>)$, for a property
$f$ of a state $x$.

\section*{Acknowledgments}

RFS was supported by NASA grant NAG 5-9563 and NSF grant AST 9819799,
for which he is grateful. This work was supported by the Norwegian
Research Council's grant 121076/420, ``Modeling of Astrophysical
Plasmas''.


\end{document}